\documentclass[english]{revtex4-1}
\usepackage[T1]{fontenc}
\usepackage[latin9]{inputenc}
\usepackage{amsmath}
\usepackage{amssymb}
\usepackage{graphicx}

\makeatletter

\providecommand{\tabularnewline}{\\}

\usepackage{babel}

\makeatother

\usepackage{babel}
\begin{document}

\title{The Linear BESS Model at the LHC}

\author{Jose Urbina}
\email{jose.urbina.avalos@gmail.com}
\author{Alfonso R. Zerwekh}
\email{alfonso.zerwekh@usm.cl}

\affiliation{Departamento de Física and Centro Científico-Tecnológico de Valparaíso,
Universidad Técnica Federico Santa María, Valparaíso, Chile}
\begin{abstract}
In this work we consider the Linear BESS model at the LHC. This model
can be seen as an adequate benchmark for exploring the phenomenological
consequences of a composite Higgs sector since its particle content
is the one we would expect in a realistic low energy description of
modern (Technicolor inspired) dynamical electroweak symmetry breaking
scenarios. Additionally, the model exhibits the property of decoupling,
producing a good ultraviolet behavior. We focus on the limits on the
masses of the new heavy vector particles imposed by direct resonance
searches, recent measurements of the decay of the Higgs boson into
two photons and the electroweak precision tests. We found that the
model is capable to accommodate the existing experimental constrains
provided that the spin-1 resonances are heavier than 3.4 TeV. 
\end{abstract}
\maketitle

\section{Introduction}

Despite its enormous experimental success, the Standard Model (SM)
offers some aspects which are not completely satisfactory from the
theoretical point of view. One of them is that not all the interactions
present in the model have their origin in the gauge principle. Indeed,
the scalar potential and the Yukawa interactions, which are central
parts of the model, are not dictated by a local symmetry. In fact,
it may be argued that this is the origin of crucial problems of the
SM like the Naturalness Problem. A very elegant solution to this criticism
is to assume that the Higgs sector and the Yukawa interactions have
their origin in a strongly interacting gauge theory. A paradigmatic
example of this framework is (extended) Technicolor, specially its
modern incarnation: Walking Technicolor. When working with strongly
interacting theories, it is usually convenient to consider effective
approaches which include only the (composite) degrees of freedom which
are relevant in the low energy limit. A very well known effective
approach to the Dynamical Electroweak Symmetry Breaking paradigm is
the BESS model \cite{BESS-1,BESS-2}. Originally, the BESS model (as
the old Technicolor idea) was Higgsless. However, the Higgs boson
was discovered and the original version is ruled out. Fortunately,
toward the end of the 1990's, a version of the BESS model which included
scalar fields was formulated. This is the so called Linear BESS (LBESS)
model \cite{LBESS-1,Casalbuoni:1997rs-1}. This model is of special
interest because its particle content (two isotriplet spin-1 resonances,
a Higgs-like scalar and two heavy scalars) is the one we would expect
in a realistic low energy description of a new strong sector responsible
for the electroweak symmetry breaking \cite{MWT,AZ-1-2}.

In this work, we study the phenomenology of this model at the LHC.
We focus on four kind of measurements in order to constrain the parameter
space of the model: the Higgs decay into a pair of photons, resonance
searches in the dijet and the dilepton spectra, and the precision
electroweak tests.

The paper is organized as follows. In section \ref{sec:A-Recall-of}
we briefly recall the main features of the LBESS model. In section
\ref{sec:Results}, we describe our simulations and results, while
in section \ref{sec:Conclusions} we state our conclusions. For the
sake of completeness we add a longer description of the model in an
Appendix.

\section{A Recall of the Linear BESS Model\label{sec:A-Recall-of}}

The basic point of view behind the BESS model attributes the origin
of the electroweak scale to a new strongly interaction sector in analogy
to the dynamical origin of $\Lambda_{\mathrm{QCD}}$ in QCD. The hypothetical
new strong interaction is supposed to be confining and at low energy
it manifests itself through composite states. It is expected that
the lightest composite particles be scalars and vector resonances.
In this section, we provide a general description of the main features
of the model. More details can be found in the Appendix and in the
original literature \cite{BESS-1,BESS-2,LBESS-1,Casalbuoni:1997rs-1}.

Following the Hidden Local Symmetry (HLS) formalism, the composite
vectors can be introduced as gauge fields of effective gauge groups.
Consequently, in the LBESS model, we start with an extended gauge
symmetry given by $SU(2)_{L}\otimes U(1)\otimes SU(2)_{L}^{\prime}\otimes SU(2)_{R}^{\prime}$
. The symmetry is broken down to $U(1)_{{\rm em}}$ in two steps as
shown in the following scheme

\[
\begin{array}{c}
SU(2)_{L}\otimes U(1)\otimes SU(2)_{L}^{\prime}\otimes SU(2)_{R}^{\prime}\\
\downarrow u\;\\
SU(2)_{{\rm weak}}\otimes U(1)_{Y}\\
\downarrow v\;\\
U(1)_{{\rm em}}
\end{array}
\]
where $u$ is the scale characterizing the breaking of the HLS and
$v$ is the usual electroweak scale. All the symmetry breaking processes
are assumed to be produced by the vacuum expectation values of (composite)
scalar fields $\left\langle \rho_{U}\right\rangle =v$ and $\left\langle \rho_{L}\right\rangle =\left\langle \rho_{R}\right\rangle =u$
. The breaking down of the symmetries produces non-diagonal mass matrices
in the gauge and scalar sector. The physical spectrum (composed by
the mass eigenstates) consists on the following fields: 
\begin{enumerate}
\item \emph{Two heavy vector triplet}: $(V_{L}^{+},V_{L}^{0},V_{L}^{-})$
and $(V_{R}^{+},V_{R}^{0},V_{R}^{-})$. These vector bosons are mainly
the gauge bosons of $SU(2)_{L}^{\prime}\otimes SU(2)_{R}^{\prime}$
with a small mixing with the gauge field of $SU(2)_{L}\otimes U(1)$.
Naturally, they have masses of the order of $g_{2}u$ where $g_{2}$is
the coupling constant associated with the groups $SU(2)_{L}^{\prime}$
and $SU(2)_{R}^{\prime}$. Because the standard fermions are assumed
to be charged only under $SU(2)_{L}$(left-handed) and $U(1)$ (left-handed
and right-handed), the heavy vectors $V_{L}^{+},V_{L}^{0},V_{L}^{-}$
and $V_{R}^{0}$ couple to the standard fermions with coupling constants
proportional to the (small) mixing angles. Notice that $V_{R}^{+}$
and $V_{R}^{-}$ do not couple to the standard fermions. 
\item \emph{The standard electroweak gauge bosons}: $W^{\pm}$, $Z$, $A$. 
\item \emph{Two heavy scalars: }$H_{L}$, $H_{R}$. These scalars are supposed
to have masses of the order of the $u$ scale. They correspond mainly
to the original $\rho_{L}$ and $\rho_{R}$ fields, respectively,
with and small mixing with $\rho_{U}$. Originally, the standard fermions
can forms Yukawa terms only with $\rho_{U}$ due to the quantum numbers
assigned to the fermions. This means that $H_{L}$ and $H_{R}$ coupling
to the standard fermions is proportional to small mixing angles. 
\item \emph{The standard-like Higgs boson: $H$} 
\end{enumerate}
The model has six free parameters, namely: the masses of the heavy
vectors ($M_{V_{L}}$and $M_{V_{R}}$), the masses of the heavy scalar
($M_{H_{L}}$and $M_{H_{R}}$), the scale of the HLS breakdown ($u$)
and a parameter governing a quartic interaction term between scalars
($f$).

In what follows, we will assume that $u$, $M_{H_{L}}$and $M_{H_{R}}$are
of the order of $3$ TeV while the masses of the heavy vectors will
be taken in the range of $2$ to $4$ TeV. The assumption of very
massive scalars beside a light Higgs-like boson is well justified
in this particular model (see equation (\ref{eq:ScalarMasses}) in
the Appendix) and has also been found to be self-consistent in a similar
effective model previously studied by our group \cite{AZ-1-2} .

\section{Results\label{sec:Results}}

\subsection{$H\rightarrow\gamma\gamma$}

\begin{figure}
\includegraphics[scale=0.7]{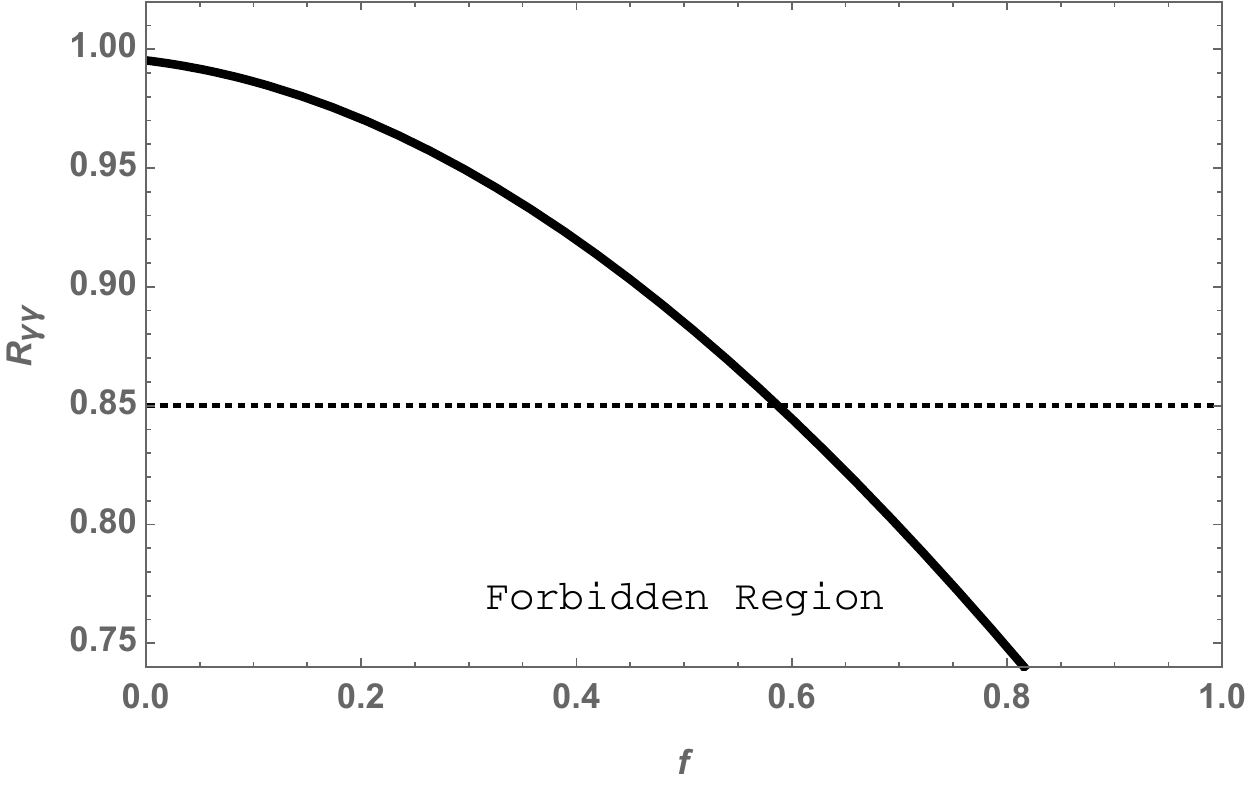}

\caption{Predicted values of $R_{\gamma\gamma}$ (continuous line), as a function
of the $f$ parameter, compared to the lower limit of the experimental
value at 1$\sigma$ (dashed line). The region above the dashed line
is allowed.}

\label{2photons} 
\end{figure}

The first process we consider is the Higgs boson decay into two photons.
This is a 1-loop process which includes the contribution of the new
charged states: in our case, the new charged vector bosons $V_{L}^{\pm}$
and $V_{R}^{\pm}$. For heavy vector bosons with moderate coupling
to the Higgs boson, it is expected that this process does not deviate
significantly from the SM \cite{AZ-1-2,AZ-2}. This is exactly our
case: the new vector bosons, as described above, are considered in
the 2-4 TeV mass range . On the other hand, the coupling between the
Higgs and $V_{R}^{\pm}$ originates from the mixing of the different
scalar fields of the model. This mixing ( and thus the referred coupling)
is controlled by the $f$ parameter of the scalar potential, which
has to be positive. In fact, we found that this parameter is the only
one sensible to the current data for this process. It is useful to
define the ratio

\begin{equation}
R_{\gamma\gamma}=\frac{\sigma\left(pp\rightarrow H\right)\Gamma\left(H\rightarrow\gamma\gamma\right)}{\sigma\left(pp\rightarrow H\right)_{\mathrm{SM}}\Gamma\left(H\rightarrow\gamma\gamma\right)_{\mathrm{SM}}}=\frac{\Gamma\left(H\rightarrow\gamma\gamma\right)}{\Gamma\left(H\rightarrow\gamma\gamma\right)_{\mathrm{SM}}}\label{eq:RAA}
\end{equation}
in order to quantify the departure of the model from the SM's predictions.
The second equality in (\ref{eq:RAA}) is due to the fact that the
Higgs production mechanism is not modified by the LBESS model. The
most recent measurement of $R_{\gamma\gamma}$ has been done by ATLAS
and results to be $R_{\gamma\gamma}=0.99\pm0.14$ \cite{Atlas-H-AA}.
In figure \ref{2photons} we show the values of $R_{\gamma\gamma}$
predicted by the LBESS model as a function of the $f$ parameter (continuous
line). For comparison purpose, we also include the lowest limit (at
$1\sigma$ ) of the experimental value. We can see that the model
is in good agreement with experiment for $f\in[0,0.6]$.

\subsection{Searching for Resonances}

We also consider the direct searches of resonances in the dijet and
the dilepton channels. In the kinematic setup we have adopted (with
very heavy non-standard scalars) only the spin-1 particle $V_{L}^{\pm},V_{L}^{0}$
(assumed to be degenerated) and $V_{R}^{0}$ can be produced in the
$s$-channel by quark--anti-quark annihilation. At this point, we
recall that the fields $V_{R}^{\pm}$ do not couple to the standard
fermions. On the other hand, in the construction we are considering,
$V_{R}$ cannot be heavier than $V_{L}$ (see equation \eqref{eq:cosphi}).
Consequently, the model predicts that two resonances should appear
in the dijet and the dilepton spectra with the lighter one corresponding
to $V_{R}^{0}$.

\subsubsection{Methodology}

\begin{figure}
\includegraphics[scale=0.4]{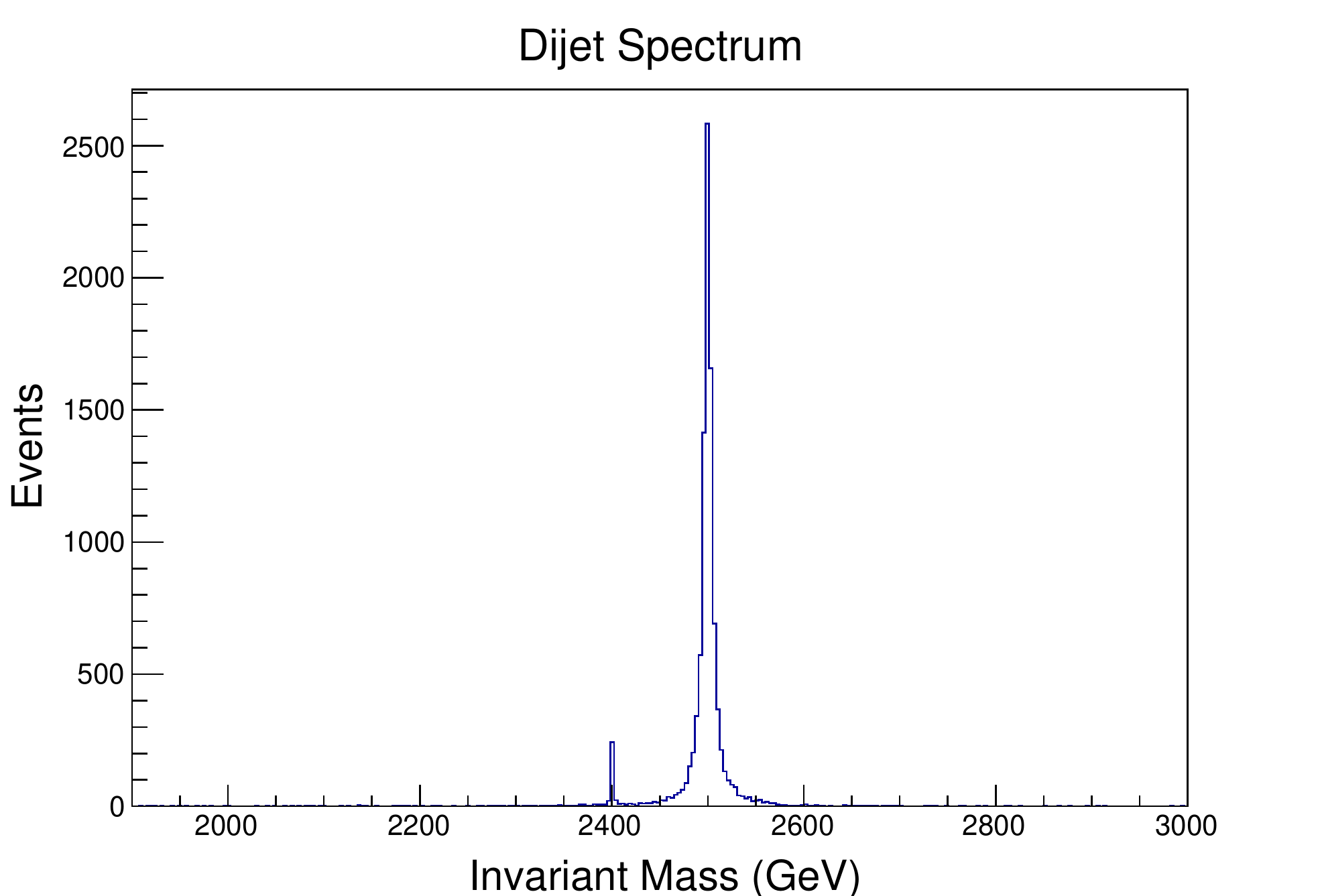}\includegraphics[scale=0.4]{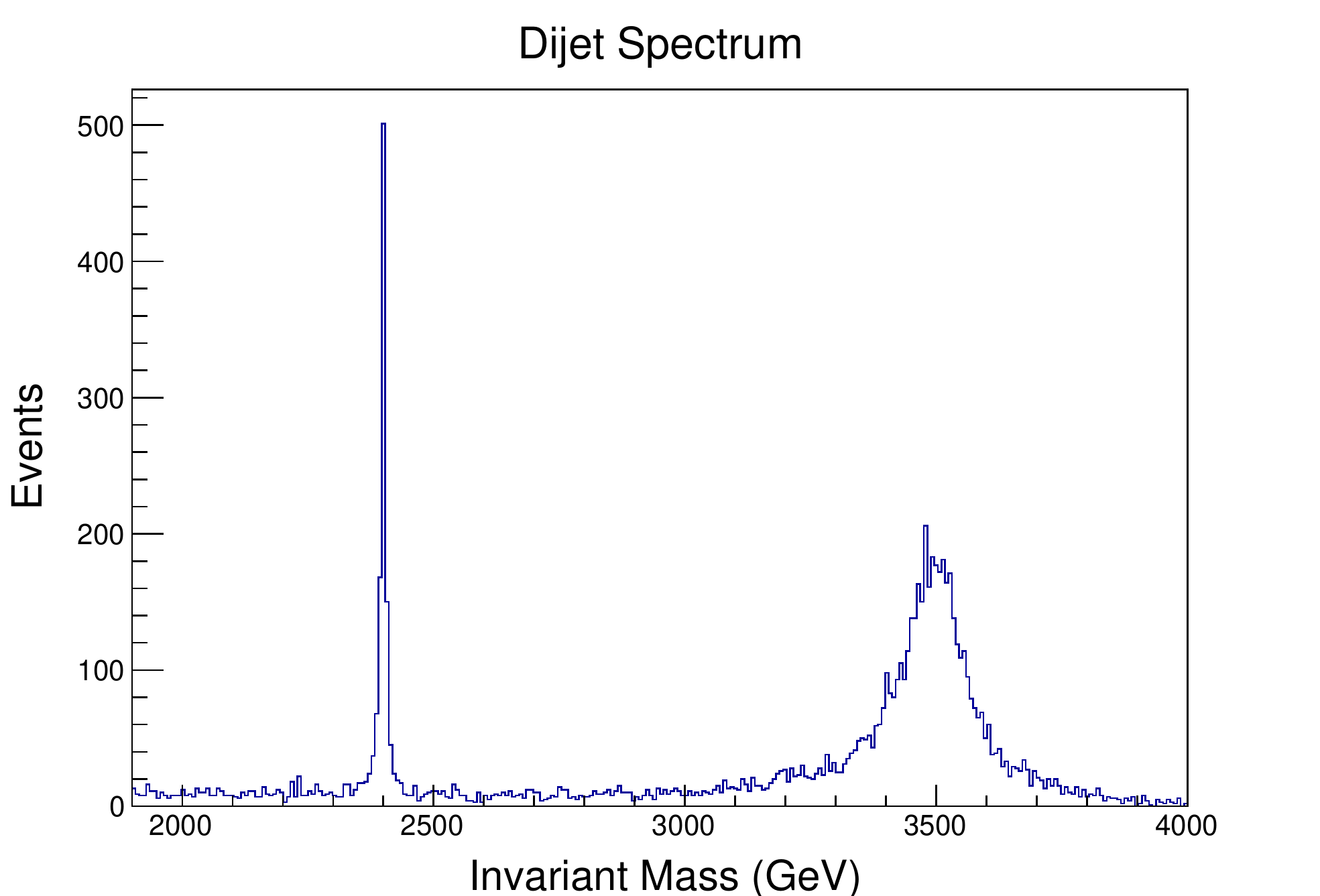}

\caption{Examples of resonances in the dijet invariant mass spectrum. In this
simulations has been taken into account only the contribution of the
non-standard sector of the LBESS model without background.}

\label{Spectra}
\end{figure}

We implemented the model in CalcHEP \cite{Calchep} using the LanHEP
package \cite{Lanhep1,Lanhep2} and generate events for dijet and
dilepton production at the 13 TeV LHC, considering only contribution
of the new particles, without background, for several values of $M_{V_{L}}$and
$M_{V_{R}}$. As an example, in Figure \ref{Spectra} we show two
dijet spectra obtained in our simulations. In order to put constrains
on the model parameter space, we count the events around each peak,
we compute a cross section for each resonance and we compare it with
the experimental upper limits for dijet or dilepton resonance cross-sections
and we only accept a pair of $M_{V_{L}}$and $M_{V_{R}}$ values if
the cross-section associated to each resonance is smaller than the
experimental limit.

\subsubsection{Dijets }

\begin{figure}
\includegraphics[scale=0.7]{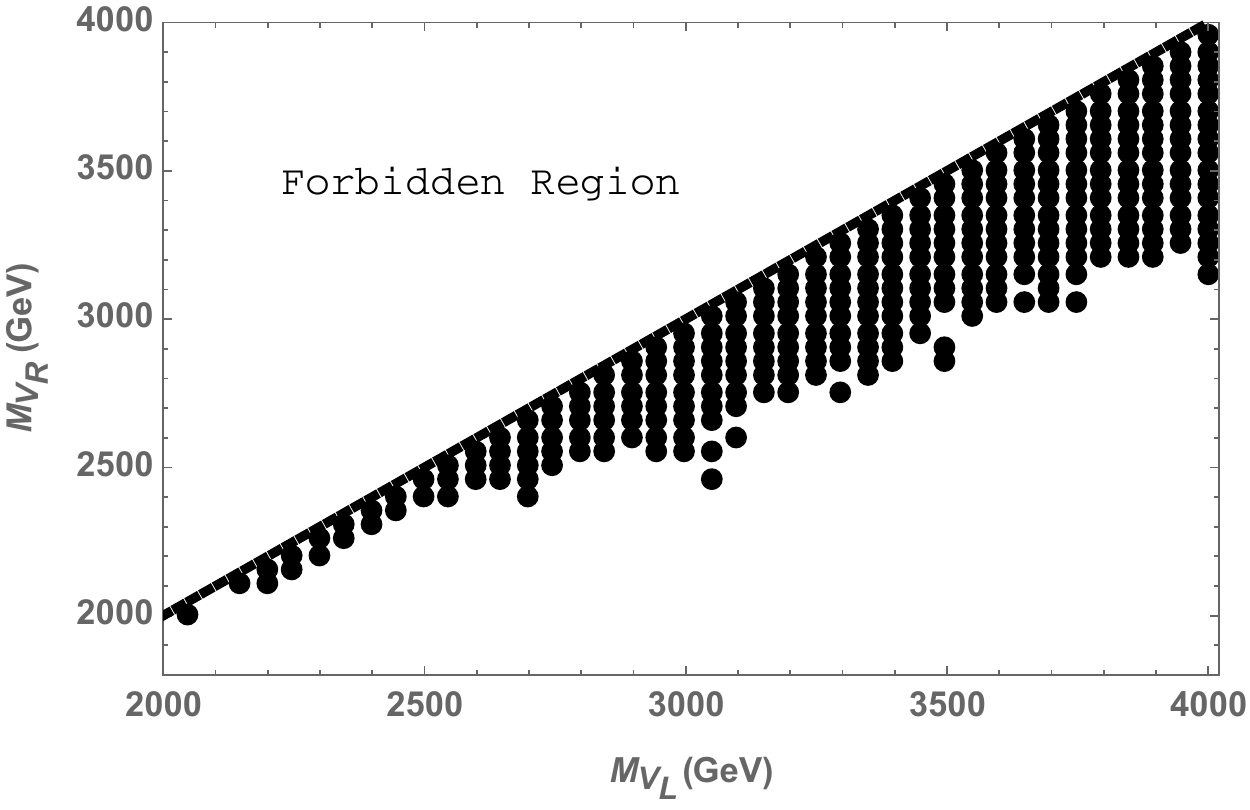}

\caption{Constrains to the $(M_{V_{L}},M_{V_{R}})$ space by direct resonance
searches in the dijets spectrum at the 13 TeV LHC. The region filled
with black circles is the allowed one.}

\label{dijet} 
\end{figure}

In the case of dijets, we use the limits provided by the ATLAS Collaboration
based on data measures at $\sqrt{s}=13$ TeV \cite{AtlasDijet}. Our
results are shown in Figure \ref{dijet}. The region filled with black
circles is the accepted zone, that is, the set of points $(M_{V_{L}},M_{V_{R}})$
which produce both resonances with a cross sections below the experimental
limit. Notice that in the quasi-degenerate case, resonances with masses
as light as 2000 or 2500 GeV are allowed. This is in agreement with
previous studies on vector resonances \cite{AZ-1-2,AZ-2,AZ-3}.

\subsubsection{Dileptons}

More stringent constrains are obtained in the dilepton channel. In
this case, we use the limits provided by the ATLAS Collaboration at
$\sqrt{s}=13$ TeV and 36.1 fb$^{-1}$\cite{AtlasDilepton}. Again,
the region filled by black circles is the accepted zone. In this case,
only resonances heavier than 3.4 TeV are allowed. This improvement
on the constrains is mainly due to the the fact that the dilepton
production is a cleaner channel than dijet production at a hadron
collider and the higher luminosity of the dilepton set of data.

\begin{figure}
\includegraphics[scale=0.7]{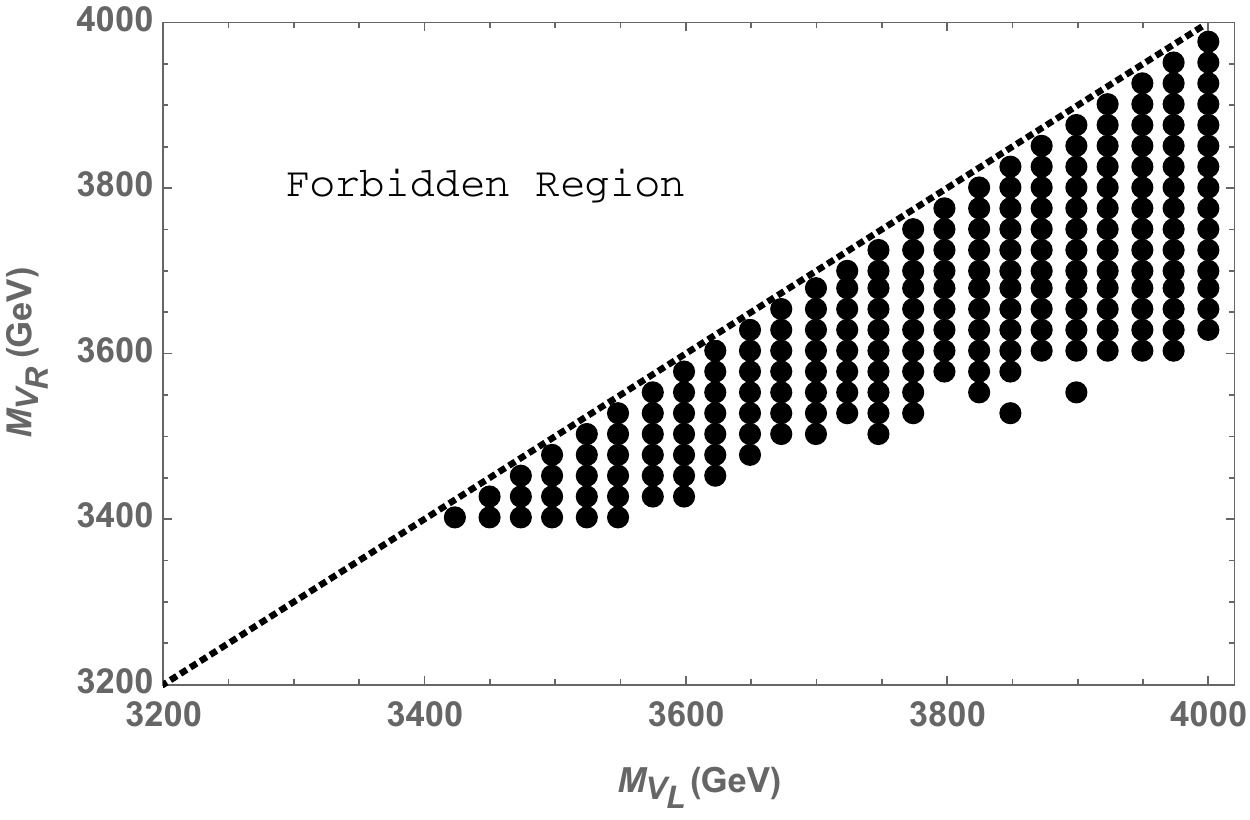}

\caption{Constrains to the $(M_{V_{L}},M_{V_{R}})$ space by direct resonance
searches in the dilepton spectrum at the 13 TeV LHC. The region filled
with black circles is the allowed one.}
\end{figure}

\subsection{Precision Tests}

An indirect way to constrain extensions of the SM is to consider the
contribution the new Physics provides to the electroweak radiative
corrections. These effects are, in general, parametrized by the well
known Peskin--Takeuchi parameters: $S$, $T$ and $U$. An equivalent
set of parameters, named $\epsilon_{1}$, $\epsilon_{2}$ and $\epsilon_{3}$,
is also used in the literature \cite{epsilon} and is related to the
former one by the following expressions:

\begin{eqnarray}
\epsilon_{1} & = & \alpha T\nonumber \\
\epsilon_{2} & = & -\frac{\alpha}{4s_{Z}^{2}}U\label{eq:epsSTU}\\
\epsilon_{3} & = & \frac{\alpha}{4s_{Z}^{2}}S\nonumber 
\end{eqnarray}
where $\alpha$ is the electromagnetic coupling constant at the scale
of $M_{Z}$ and $s_{Z}^{2}$ is the $\sin^{2}\theta_{W}$ in the $\overline{MS}$
scheme at the same scale. In Ref. \cite{Casalbuoni:1997rs-1}, the
following tree level expressions are provided for the $\epsilon_{i}$
parameters in the context of the LBESS model:

\begin{eqnarray}
\epsilon_{1} & = & -rs_{\varphi}^{2}\frac{c_{\theta}^{4}+s_{\theta}^{4}}{c_{\theta}^{4}}\nonumber \\
\epsilon_{2} & = & -rs_{\varphi}^{2}\label{eq:eps}\\
\epsilon_{3} & = & -\frac{rs_{\varphi}^{2}}{c_{\theta}^{2}}\nonumber 
\end{eqnarray}
where $c_{\theta}=\cos\theta_{W}$, $s_{\theta}=\sin\theta_{W}$,
while $r$ and $s_{\varphi}$ can be expressed in terms of $M_{V_{L}}$and
$M_{V_{R}}$ as shown in the Appendix.

\begin{figure}
\includegraphics[scale=0.4]{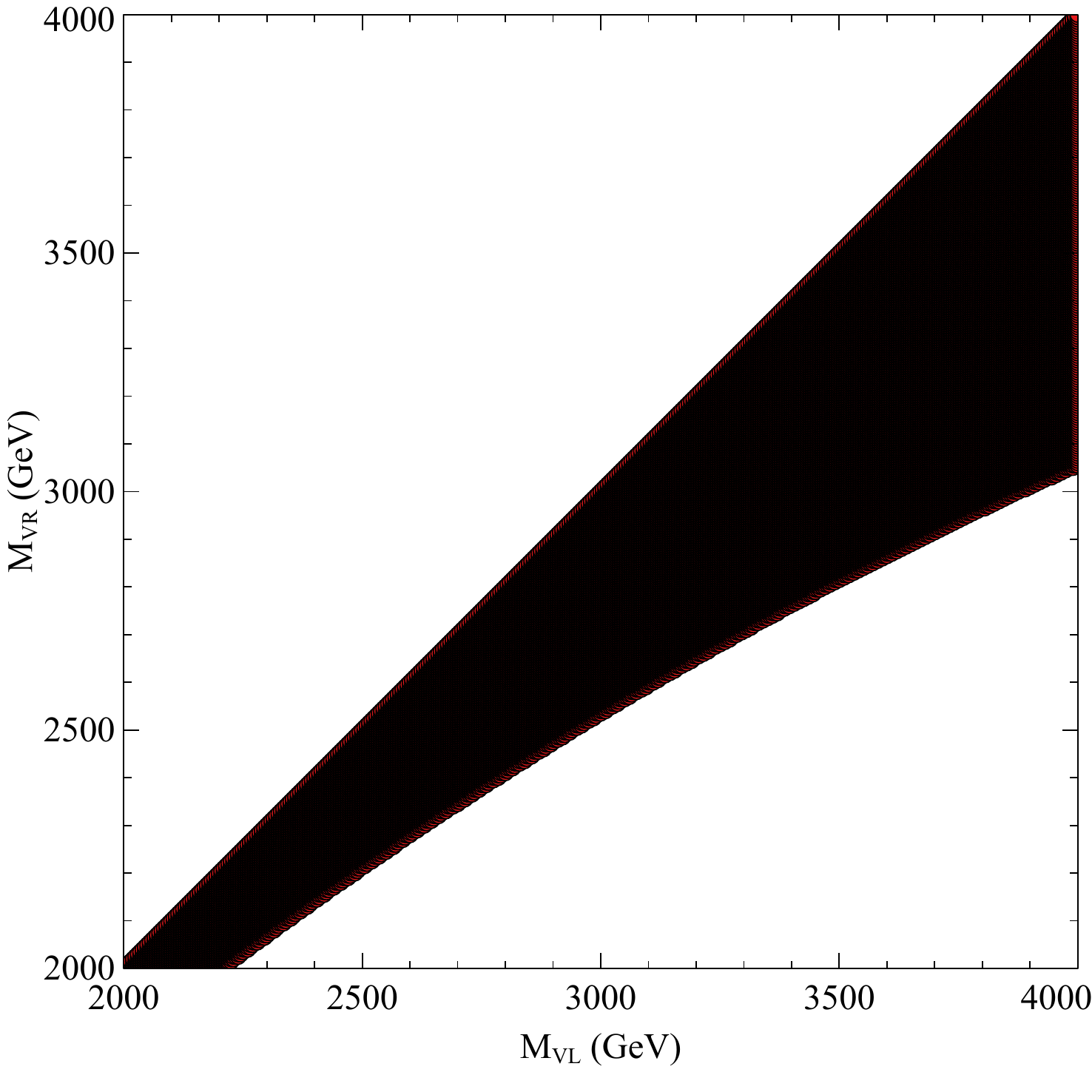}

\caption{Constrains to the $(M_{V_{L}},M_{V_{R}})$ space by the electroweak
precision tests. The black region is the allowed one.}

\label{fig:EWtests}
\end{figure}

We use the values $S=0.05\pm0.10$, $T=0.08\pm0.12$ and $U=0.02\pm0.10$
found in \cite{PDG}, and the expressions above to select the combination
of $M_{V_{L}}$and $M_{V_{R}}$ which reproduce simultaneously the
experimental values of $\epsilon_{1}$, $\epsilon_{2}$ and $\epsilon_{3}$
(using equations (\ref{eq:eps})) within $1\sigma$. The result is
shown in Fig. \ref{fig:EWtests}. The black region is the zone allowed
by the precision variables. As we can see, the restrictions are not
as stringent as the ones obtained using dilepton data.

\section{Conclusions\label{sec:Conclusions}}

We have studied the LBESS model in the context of recent data obtained
at the 13 TeV LHC in the regime where the non-standard scalars are
heavy with masses of the order of 3 TeV. We found that the value the
$f$ parameter of the scalar potential has to belong to the interval
$[0,0.6].$ Additionally, we have found that the model is consistent
with current experimental data provided that the spin-1 resonances
are heavier than 3.4 TeV. This limit is higher than the ones obtained
in previous studies of models with vector resonances. The main restrictions
come from recent searches of resonances in the dilepton spectrum.

\section*{Appendix\label{sec:Appendix}}

The LBESS model is based on the global symmetry $SU(2)_{L}\otimes SU(2)_{R}\otimes SU(2)_{L}^{\prime}\otimes SU(2)_{R}^{\prime}$
from which the subgroup $SU(2)_{L}\otimes U(1)\otimes SU(2)_{L}^{\prime}\otimes SU(2)_{R}^{\prime}$
is made local. The matter content consist of three scalars ($\Phi_{{\rm U}}$,
$\Phi_{{\rm L}}$ and $\Phi_{{\rm R}}$) which we are supposed to
be composite and the standard fermions (plus a right-handed neutrino).
These fields transform under the global symmetry following the representation
assignment showed in table \ref{table1} (scalars) and \ref{table3}
(fermions). The representation assignment for fermions under the local
symmetry is shown in table \ref{table2}.

\begin{table}[htbp]
\centering{}%
\begin{tabular}{|c|c|c|c|c|}
\hline 
Fields  & $SU(2)_{L}$  & $SU(2)_{R}$  & $SU(2)_{L}^{\prime}$  & $SU(2)_{R}$ \tabularnewline
\hline 
\hline 
$\Phi_{{\rm U}}$  & ${\bf 2}$  & ${\bf \bar{2}}$  & $1$  & $1$ \tabularnewline
\hline 
$\Phi_{{\rm L}}$  & ${\bf 2}$  & $1$  & ${\bf 2}$  & $1$ \tabularnewline
\hline 
$\Phi_{{\rm R}}$  & $1$  & ${\bf 2}$  & $1$  & ${\bf 2}$ \tabularnewline
\hline 
\end{tabular}\caption{Representations of the scalar fields under the full global group.}
\label{table1} 
\end{table}

\begin{table}[htbp]
\centering{}%
\begin{tabular}{|c|c|c|c|c|}
\hline 
Fermion  & $SU(2)_{L}$  & $SU(2)_{R}$  & $SU(2)_{L}^{\prime}$  & $SU(2)_{R}^{\prime}$ \tabularnewline
\hline 
\hline 
$\Psi_{iL}$  & ${\bf 2}$  & $1$  & $1$  & $1$ \tabularnewline
\hline 
$\Psi_{iR}$  & $1$  & ${\bf 2}$  & $1$  & $1$ \tabularnewline
\hline 
\end{tabular}\caption{Representations of the fermion fields under the full global group.}
\label{table3} 
\end{table}

\begin{table}[htbp]
\centering{}%
\begin{tabular}{|c|c|c|c|c|}
\hline 
Fermion  & $SU(2)_{L}$  & $U(1)$  & $SU(2)_{L}^{\prime}$  & $SU(2)_{R}^{\prime}$ \tabularnewline
\hline 
\hline 
$\Psi_{iL}$  & ${\bf 2}$  & $Y^{\prime}$  & $1$  & $1$ \tabularnewline
\hline 
$\Psi_{iR}$  & $1$  & $Y^{\prime}$  & $1$  & $1$ \tabularnewline
\hline 
\end{tabular}\caption{ Representations of the fermion fields under the full gauge group.
Here $Y^{\prime}=B-L$ with $B$ and $L$ being the baryon and lepton
numbers.}
\label{table2} 
\end{table}

The full Lagrangian of model can be written down as:

\begin{equation}
\begin{aligned}{\cal L} & =\frac{1}{2}{\rm Tr}[F_{\mu\nu}(\tilde{W})F^{\mu\nu}(\tilde{W})]+\frac{1}{2}{\rm Tr}[F_{\mu\nu}(\tilde{B})F^{\mu\nu}(\tilde{B})]+\frac{1}{2}{\rm Tr}[F_{\mu\nu}(A_{L})F^{\mu\nu}(A_{L})]+\frac{1}{2}{\rm Tr}[F_{\mu\nu}(A_{R})F^{\mu\nu}(A_{R})]\\
 & +\frac{1}{4}\left[{\rm Tr}[(D_{\mu}\Phi_{{\rm U}})^{\dagger}(D^{\mu}\Phi_{{\rm U}})]+{\rm Tr}[(D_{\mu}\Phi_{{\rm L}})^{\dagger}(D^{\mu}\Phi_{{\rm L}})]+{\rm Tr}[(D_{\mu}\Phi_{{\rm R}})^{\dagger}(D^{\mu}\Phi_{{\rm R}})]\right]\\
 & -\mu^{2}({\rm Tr}[\Phi_{{\rm L}}^{\dagger}\Phi_{{\rm L}}]+{\rm Tr}[\Phi_{{\rm R}}^{\dagger}\Phi_{{\rm R}}])-\frac{\lambda}{4}([{\rm Tr}(\Phi_{{\rm L}}^{\dagger}\Phi_{{\rm L}})]^{2}+[{\rm Tr}(\Phi_{{\rm R}}^{\dagger}\Phi_{{\rm R}})]^{2})-m^{2}{\rm Tr}(\Phi_{{\rm U}}^{\dagger}\Phi_{{\rm U}})-\frac{h}{4}[{\rm Tr}(\Phi_{{\rm U}}^{\dagger}\Phi_{{\rm U}})]^{2}\\
 & -\frac{f_{3}}{2}{\rm Tr}(\Phi_{{\rm L}}^{\dagger}\Phi_{{\rm L}}){\rm Tr}(\Phi_{{\rm R}}^{\dagger}\Phi_{{\rm R}})-\frac{f}{2}[{\rm Tr}(\Phi_{{\rm L}}^{\dagger}\Phi_{{\rm L}}){\rm Tr}(\Phi_{{\rm U}}^{\dagger}\Phi_{{\rm U}})+{\rm Tr}(\Phi_{{\rm R}}^{\dagger}\Phi_{{\rm R}}){\rm Tr}(\Phi_{{\rm U}}^{\dagger}\Phi_{{\rm U}})]\\
 & +i\bar{\Psi}_{iL}\gamma^{\mu}D_{\mu}\Psi_{iL}+i\bar{\Psi}_{iR}\gamma^{\mu}D_{\mu}\Psi_{iR}+\bar{\Psi}_{iL}\Phi_{{\rm U}}\Upsilon\Psi_{jR}+{\rm h.c.}\label{lagrangian}
\end{aligned}
\end{equation}
where $\tilde{W}_{\mu},\tilde{B}_{\mu},A_{L\mu}$ and $A_{R\mu}$
are the gauge bosons associated to $SU(2)_{L}$, $U(1)$, $SU(2)_{L}^{\prime}$
and $SU(2)_{R}^{\prime}$, respectively and $\Upsilon$ is a matrix
containing the Yukawa coupling constants. On the other hand, the strength
tensors a defines as usual:

\begin{equation}
\begin{aligned}F_{\mu\nu}(\tilde{W}) & =\partial_{\mu}\tilde{W}_{\nu}-\partial_{\nu}\tilde{W}_{\mu}+g_{0}[\tilde{W}_{\mu},\tilde{W}_{\nu}]\\
F_{\mu\nu}(\tilde{B}) & =\partial_{\mu}\tilde{B}_{\nu}-\partial_{\nu}\tilde{B}_{\mu}\\
F_{\mu\nu}(A_{L}) & =\partial_{\mu}A_{L\nu}-\partial_{\nu}A_{L\mu}+g_{2}[A_{L\mu},A_{L\nu}]\\
F_{\mu\nu}(A_{R}) & =\partial_{\mu}A_{R\nu}-\partial_{\nu}A_{R\mu}+g_{2}[A_{R\mu},A_{R\nu}]\label{kinecterm}
\end{aligned}
\end{equation}

The covariant derivatives in the kinetic terms for the scalars and
fermions in equation (\ref{lagrangian}), are given by:

\begin{equation}
\begin{aligned}D_{\mu}\Phi_{{\rm L}} & =\partial_{\mu}\Phi_{{\rm L}}+ig_{0}\frac{\tau^{a}}{2}\tilde{W}_{\mu}^{a}\Phi_{{\rm L}}-ig_{2}\Phi_{{\rm L}}\frac{\tau^{a}}{2}A_{L\mu}^{a}\\
D_{\mu}\Phi_{{\rm R}} & =\partial_{\mu}\Phi_{{\rm R}}+ig_{1}\frac{\tau_{3}}{2}\tilde{B}_{\mu}\Phi_{{\rm R}}-ig_{2}\Phi_{{\rm R}}\frac{\tau^{a}}{2}A_{R\mu}^{a}\\
D_{\mu}\Phi_{{\rm U}} & =\partial_{\mu}\Phi_{{\rm U}}+ig_{0}\frac{\tau^{a}}{2}\tilde{W}_{\mu}^{a}\Phi_{{\rm U}}-ig_{1}\Phi_{{\rm U}}\frac{\tau_{3}}{2}\tilde{B}_{\mu}
\end{aligned}
\label{scalarderivate}
\end{equation}

\begin{equation}
\begin{aligned}D_{\mu}\Psi_{iL} & =(\partial_{\mu}+ig_{0}\tilde{W}_{\mu}^{a}\frac{\tau^{a}}{2}+\frac{i}{2}g_{1}Y^{\prime}\tilde{B}_{\mu})\Psi_{iL}\\
D_{\mu}\Psi_{iR} & =(\partial_{\mu}+ig_{1}\tilde{B}_{\mu}\frac{\tau^{3}}{2}+\frac{i}{2}g_{1}Y^{\prime}\tilde{B}_{\mu})\Psi_{iR}
\end{aligned}
\end{equation}
where $g_{0}$, $g_{1}$ and $g_{2}$ are the coupling constants associated
to the groups $SU(2)_{L}$, $U(1)$ and $SU(2)_{L}^{\prime}\otimes SU(2)_{R}^{\prime}$
respectively.

For the aim of simplicity, it has been assumed an interchange symmetry
between $\Phi_{{\rm L}}$ and $\Phi_{{\rm R}}$ in the potential and
in the kinetic terms. Writing the scalars in ``polar parametrization'',
\emph{i.e. }$\Phi_{{\rm L}}=\rho_{L}L$, $\Phi_{{\rm R}}=\rho_{R}R$
and $\Phi_{{\rm U}}=\rho_{U}U$ where $L$, $R$ and $U$ are unitary
matrices, the potential gets a simpler form:

\begin{equation}
\begin{aligned}V(\rho_{U},\rho_{L},\rho_{R}) & =2\mu^{2}\left[(\rho_{L}+u)^{2}+(\rho_{R}+u)^{2}\right]+\lambda\left[(\rho_{L}+u)^{4}+(\rho_{R}+u)^{4}\right]\\
 & +2m^{2}(\rho_{U}+v)^{2}+h(\rho_{U}+v)^{4}+2f_{3}(\rho_{L}+u)^{2}(\rho_{R}+u)^{2}\\
 & +2f(\rho_{U}+v)^{2}\left[(\rho_{L}+u)^{2}+(\rho_{R}+u)^{2}\right]\label{potential}
\end{aligned}
\end{equation}

The scalar fields acquire a vacuum expectation value (vev): $\left\langle \rho_{U}\right\rangle =v$
and $\left\langle \rho_{L}\right\rangle =\left\langle \rho_{R}\right\rangle =u$
which spontaneously break the original gauge symmetry down to $U(1)_{{\rm em}}$.

In the true vacuum, nontrivial mass matrices appear in the scalar
and the vector sector implying that the mass eigenstate are different
from the ``flavor'' ones. In the case of scalar the relationship
between flavor and mass eigenvectors is given, in in the limit $u\gg v$,
by:

\begin{eqnarray}
\begin{bmatrix}\rho_{L}\\
\rho_{R}\\
\rho_{U}
\end{bmatrix} & = & \begin{bmatrix}\frac{1}{\sqrt{2}} & \frac{1}{\sqrt{2}}-\frac{q^{2}}{s_{\varphi}^{2}}r & -\frac{q}{s_{\varphi}}\sqrt{r}\\
-\frac{1}{\sqrt{2}} & \frac{1}{\sqrt{2}}(1-\frac{q^{2}}{s_{\varphi}^{2}}r) & -\frac{q}{s_{\varphi}}\sqrt{r}\\
0 & \frac{q}{s_{\varphi}}\sqrt{2r} & 1-\frac{q^{2}}{s_{\varphi}^{2}}r
\end{bmatrix}\begin{bmatrix}H_{L}\\
H_{R}\\
H
\end{bmatrix}\label{scalarfield}
\end{eqnarray}
where the variables $r$ and $q$ are

\begin{equation}
r=\frac{v^{2}}{u^{2}}\frac{g^{2}}{g_{2}^{2}}~~~~~~~~q=\frac{f}{f_{3}+\lambda}\label{rq}
\end{equation}
,

\[
s_{\varphi}=\sin(\varphi)=\frac{g_{0}}{\sqrt{g_{0}^{2}+g_{2}^{2}}}
\]
and, $H_{L}$ and $H_{R}$ are the physical heavy scalar while $H$
denotes the standard-like Higgs boson.

The mass eigenvalues for the scalar fields, in the limit $u\gg v$,
are given by:

\begin{equation}
\begin{aligned}M_{H}^{2} & =8v^{2}(h-2\frac{f^{2}}{f_{3}+\lambda})\\
M_{H_{L}}^{2} & =8u^{2}(\lambda-{f_{3}})\\
M_{H_{R}}^{2} & =8u^{2}(\lambda+{f_{3}})
\end{aligned}
\label{eq:ScalarMasses}
\end{equation}

Similarly, in the vector sector, the relationship between flavor and
mass eigenvectors is given by :

\begin{equation}
\begin{bmatrix}\tilde{W}^{\pm}\\
A_{L}^{\pm}
\end{bmatrix}=\begin{bmatrix}c_{\varphi}(1-s_{\varphi}^{2}r) & -s_{\varphi}(1+c_{\varphi}^{2}r)\\
s_{\varphi}(1+c_{\varphi}^{2}r) & c_{\varphi}(1-s_{\varphi}^{2}r)
\end{bmatrix}\begin{bmatrix}W^{\pm}\\
V_{L}^{\pm}
\end{bmatrix}
\end{equation}

\begin{equation}
\begin{bmatrix}\tilde{W}_{3}\\
\tilde{B}\\
A_{L}^{3}\\
A_{R}^{3}
\end{bmatrix}=\begin{bmatrix}c_{\varphi}s_{\theta} & c_{\varphi}(c_{\theta}-\frac{s_{\varphi}^{2}}{c_{\theta}}r) & -s_{\varphi}(1+c_{\varphi}^{2}r) & \frac{c_{\varphi}s_{\varphi}s_{\theta}^{4}\sqrt{P}}{c_{\theta}^{3}(1-2c_{\theta}^{2})}r\\
\sqrt{P} & -\frac{s_{\theta}}{c_{\theta}}\sqrt{P}(1-\frac{s_{\varphi}^{2}s_{\theta}^{2}}{c_{\theta}^{4}}r) & -\frac{c_{\varphi}s_{\varphi}s_{\theta}\sqrt{P}}{1-2c_{\theta}^{2}}r & -\frac{s_{\varphi}s_{\theta}}{c_{\theta}}(1+\frac{s_{\theta}^{2}P}{c_{\theta}^{4}}r)\\
s_{\varphi}s_{\theta} & s_{\varphi}c_{\theta}(1+\frac{c_{\varphi}^{2}}{c_{\theta}^{2}}r) & c_{\varphi}(1-s_{\varphi}^{2}r) & -\frac{s_{\theta}^{2}P^{3/2}}{c_{\theta}^{3}(1-2c_{\theta}^{2})}r\\
s_{\varphi}s_{\theta} & -\frac{s_{\varphi}s_{\theta}^{2}}{c_{\theta}}(1+\frac{P}{c_{\theta}^{4}}r) & \frac{c_{\varphi}^{3}s_{\theta}^{2}}{1-2c_{\theta}^{2}}r & \frac{\sqrt{P}}{c}_{\theta}(1-\frac{s_{\varphi}^{2}s_{\theta}^{4}}{c_{\theta}^{4}}r)
\end{bmatrix}\begin{bmatrix}A\\
Z\\
V_{L}^{0}\\
V_{R}^{0}
\end{bmatrix}
\end{equation}

Where, as usual $A$,$W^{\pm}$and $Z$ represent the photon and the
standard weak gauge bosons while $V_{L}^{\pm,0}$ and $V_{R}^{\pm,0}$
are the physical heavy vector bosons and $\theta$ represents the
Weinberg angle.

The mass eigenvalues of the vector sector are:

\begin{equation}
\begin{aligned}M_{A}^{2} & =0\\
M_{Z}^{2} & =\frac{v^{2}}{4}\frac{g^{2}}{c_{\theta}^{2}}(1-rs_{\varphi}^{2}\frac{1-2c_{\theta}^{2}+2c_{\theta}^{4}}{c_{\theta}^{4}}+\cdots)\\
M_{V_{L}^{0}}^{2} & =\frac{v^{2}}{4}g^{2}(\frac{1}{rc_{\varphi}^{2}}+\frac{s_{\varphi}^{2}}{c_{\varphi}^{2}}-rs_{\varphi}^{2}\frac{c_{\theta}^{2}}{1-2c_{\theta}^{2}}+\cdots)\\
M_{V_{R}^{0}}^{2} & =\frac{v^{2}}{4}\frac{g^{2}}{c_{\theta}^{2}}(\frac{1}{r}\frac{c_{\theta}^{4}}{P}+\frac{s_{\varphi}^{2}s_{\theta}^{4}}{P}+r\frac{s_{\varphi}^{2}s_{\theta}^{8}}{c_{\theta}^{4}(1-2c_{\theta}^{2})}+\cdots)\\
M_{V_{R}}^{2} & =\frac{1}{4}g_{2}^{2}u^{2}\\
M_{W}^{2} & =\frac{v^{2}}{4}g^{2}(1-rs_{\varphi}^{2}+\cdots)\\
M_{V_{L}}^{2} & =\frac{v^{2}}{4}g^{2}(\frac{1}{r}\frac{1}{c_{\varphi}^{2}}+\frac{s_{\varphi}^{2}}{c_{\varphi}^{2}}+rs_{\varphi}^{2}+\cdots)
\end{aligned}
\end{equation}

From the equations above it is easy to see that, when $u\gg v$, it
is possible to write the $r$ parameter and $c_{\varphi}$in the convenient
form:

\begin{equation}
r\thickapprox\frac{M_{W}^{2}}{M_{V_{R}}^{2}}
\end{equation}

\begin{equation}
c_{\varphi}\approx\frac{M_{V_{R}}}{M_{V_{L}}}\label{eq:cosphi}
\end{equation}

Notice that due to the representation assignments, the fermions only
couple to the gauge bosons $\tilde{W}$ and $B$ so their coupling
to the heavy vector mass eigenstates arise only through mixing terms.

In this model, as shown in the Lagrangian, only Yukawa terms involve
the fermions and the scalar field $\Phi_{{\rm U}}$ are allowed by
the global symmetry. The Yukawa Lagrangian can be expanded as follows:

\begin{equation}
{\cal L}_{Y}=\sum_{i.j}^{3}[y_{ij}^{d}(\bar{L}_{i}^{q}\Phi_{U})R_{j}^{d}+y_{ij}^{u}(\bar{L}_{i}^{q}\tilde{\Phi_{U}})R_{j}^{u}+y_{ij}^{l_{d}}(\bar{L}_{i}^{l}\Phi_{U})R_{j}^{l_{d}}+y_{ij}^{l_{u}}(\bar{L}_{i}^{l}\Phi_{U})R_{j}^{l_{u}}+{\rm h.c.}]
\end{equation}
where the components of the scalar field are:

\begin{equation}
\Phi_{U}=\begin{bmatrix}iw^{+}\\
\frac{(v+\rho_{{\rm U}})+iz}{\sqrt{2}}.
\end{bmatrix}
\end{equation}
and

\begin{equation}
\rho_{U}=(1-\frac{q^{2}}{s_{\varphi}^{2}}r)H+\frac{q}{s_{\varphi}}\sqrt{2r}H_{R}.
\end{equation}

It should also be noted that $\cos\varphi$ can be approximated to:

It is important to note that the parameters of the scalar potential
have theoretical restrictions, which are described below:

\begin{equation}
\begin{aligned}\mu^{2} & <0~~~~~~~~m^{2}<0~~~~~~~~f>0\\
\\
\lambda-f_{3} & >0~~~~~~h>f\frac{m^{2}}{\mu^{2}}~~~~~~\lambda+f_{3}>2f\frac{\mu^{2}}{m^{2}}\label{restrteo}
\end{aligned}
\end{equation}

The restrictions for $\mu^{2}$ and $m^{2}$ are imposed so that the
potential acquires a vacuum expectation value other than zero. On
the other hand the restriction for $f$ comes from the decoupling
of the model to the standard model with Higgs. The remaining restrictions
in equation (\ref{restrteo}) derive from the positivity of the mass
spectrum of the scalar fields.

Finally, we offer in table \ref{tab:summary} a summary of the free
parameters of the model.

\begin{table}
\begin{centering}
\begin{tabular}{|c|c|}
\hline 
Parameters  & Meaning \tabularnewline
\hline 
$u$  & Scale at which the extended symmetry breaks down \tabularnewline
\hline 
$f$  & Parameter of quartic interactions in the potential \tabularnewline
\hline 
$M_{H_{L}}$  & Mass of the heavy scalar $H_{L}$ \tabularnewline
\hline 
$M_{H_{R}}$  & Mass of the heavy scalar$H_{R}$ \tabularnewline
\hline 
$M_{V_{L}}$  & Mass of the heavy vector $V_{L}^{\pm},~~V_{L}^{0}$ \tabularnewline
\hline 
$M_{V_{R}}$  & Mass of the heavy vector $V_{R}^{\pm},~~V_{R}^{0}$ \tabularnewline
\hline 
\end{tabular}
\par\end{centering}
\caption{Summary of the model's free parameters }

\label{tab:summary}
\end{table}

\section*{Acknowledgements}
The authors want to thank to Bastián Díaz and Felipe Rojas for
useful discussions.

This work was supported in part by Conicyt (Chile) grants PIA/ACT-1406
and PIA/Basal FB0821, and by Fondecyt (Chile) grant 1160423.


\begin{thebibliography}{10}
\bibitem{BESS-1}Casalbuoni R., De Curtis S., Dominici D. and Gatto
R., Phys. Lett. B, \textbf{155} (1985) 95

\bibitem{BESS-2}Casalbuoni R., De Curtis S., Dominici D. and Gatto
R., Nucl. Phys. B, \textbf{282} (1987) 235

\bibitem{LBESS-1}R. Casalbuoni, S. De Curtis, D. Dominici and M.
Grazzini, ``An Extension of the electroweak model with decoupling
at low-energy,'' Phys. Lett. B \textbf{388}, 112 (1996) doi:10.1016/0370-2693(96)01129-X
{[}hep-ph/9607276{]}.

\bibitem{Casalbuoni:1997rs-1}R. Casalbuoni, S. De Curtis, D. Dominici
and M. Grazzini, ``New vector bosons in the electroweak sector: A
Renormalizable model with decoupling,'' Phys. Rev. D\textbf{56},
5731 (1997) {[}hep-ph/9704229{]}.

\bibitem{MWT}R. Foadi, M. T. Frandsen, T. A. Ryttov and F. Sannino,
``Minimal Walking Technicolor: Set Up for Collider Physics,'' Phys.
Rev. D \textbf{76}, 055005 (2007) doi:10.1103/PhysRevD.76.055005 {[}arXiv:0706.1696
{[}hep-ph{]}{]}.

\bibitem{AZ-1-2}A. E. Carcamo Hernandez, C. O. Dib and A. R. Zerwekh,
``The Effect of Composite Resonances on Higgs decay into two photons,''
Eur. Phys. J. C , 2822 (2014) doi:10.1140/epjc/s10052-014-2822-6 {[}arXiv:1304.0286
{[}hep-ph{]}{]}.

\bibitem{AZ-2}A. E. Cárcamo Hernández, B. Díaz Sáez, C. O. Dib and
A. Zerwekh, ``Constraints on vector resonances from a strong Higgs
sector,'' Phys. Rev. D \textbf{96} (2017) no.11, 115027 doi:10.1103/PhysRevD.96.115027
{[}arXiv:1707.05195 {[}hep-ph{]}{]}.

\bibitem{Atlas-H-AA}M. Aaboud \emph{et al.} {[}ATLAS Collaboration{]},
``Measurements of Higgs boson properties in the diphoton decay channel
with 36 fb$^{-1}$ of $pp$ collision data at $\sqrt{s}=13$ TeV with
the ATLAS detector,'' arXiv:1802.04146 {[}hep-ex{]}.

\bibitem{Calchep}A. Belyaev, N. D. Christensen and A. Pukhov, ``CalcHEP
3.4 for collider physics within and beyond the Standard Model,''
Comput. Phys. Commun. \textbf{184}, 1729 (2013) doi:10.1016/j.cpc.2013.01.014
{[}arXiv:1207.6082 {[}hep-ph{]}{]}.

\bibitem{Lanhep1}A. Semenov, ``LanHEP: A Package for the automatic
generation of Feynman rules in field theory. Version 3.0,'' Comput.
Phys. Commun. \textbf{180}, 431 (2009) doi:10.1016/j.cpc.2008.10.012
{[}arXiv:0805.0555 {[}hep-ph{]}{]}.

\bibitem{Lanhep2}A. Semenov, ``LanHEP --- A package for automatic
generation of Feynman rules from the Lagrangian. Version 3.2,'' Comput.
Phys. Commun. \textbf{201}, 167 (2016) doi:10.1016/j.cpc.2016.01.003
{[}arXiv:1412.5016 {[}physics.comp-ph{]}{]}.

\bibitem{AtlasDijet}The ATLAS collaboration {[}ATLAS Collaboration{]},
``Search for New Phenomena in Dijet Events with the ATLAS Detector
at $\sqrt{s}=13$ TeV with 2015 and 2016 data,'' ATLAS-CONF-2016-069.

\bibitem{AZ-3}O. Castillo-Felisola, C. Corral, M. González, G. Moreno,
N. A. Neill, F. Rojas, J. Zamora and A. R. Zerwekh, ``Higgs Boson
Phenomenology in a Simple Model with Vector Resonances,'' Eur. Phys.
J. C \textbf{73}, no. 12, 2669 (2013) doi:10.1140/epjc/s10052-013-2669-2
{[}arXiv:1308.1825 {[}hep-ph{]}{]}.

\bibitem{AtlasDilepton}The ATLAS collaboration {[}ATLAS Collaboration{]},
``Search for new high-mass phenomena in the dilepton final state
using 36.1 fb$^{-1}$ of proton-proton collision data at $\sqrt{s}=13$
TeV with the ATLAS detector,'' JHEP \textbf{1710}, 182 (2017) doi:10.1007/JHEP10(2017)182
{[}arXiv:1707.02424 {[}hep-ex{]}{]}.

\bibitem{epsilon}G. Altarelli and R. Barbieri, Phys. Lett. \textbf{B253}
, 161 (1991)

\bibitem{PDG}C. Patrignani et al. (Particle Data Group), Chin. Phys.
C, 40, 100001 (2016) and 2017 update. 
\end{thebibliography}
\end{document}